# Generic Description of Well-Scoped, Well-Typed Syntaxes

Gergő Érdi

gergo@erdi.hu

We adapt the technique of type-generic programming via descriptions pointing into a universe to the domain of typed languages with binders and variables, implementing a notion of *syntax-generic programming* in a dependently typed programming language.

We present an Agda library implementation of type-preserving renaming and substitution (including proofs about their behaviour) "once and for all" over all applicable languages using our technique.

## 1  Introduction

Type theory-based tools like Agda[22] or Coq[6] are frequently used to study formal languages, which are often typed, and involve structures that bind and that reference names. All these aspects: syntax, scoping rules, and typing rules, need to be modeled, so that we can prove properties and implement transformations in the metalanguage of our choice.

The usual approach to this modeling work is to create bespoke datatypes in the metalanguage for each object language, and write, from scratch, all proofs and transformations, as needed, in terms of those datatypes. While there are standard techniques for creating the right representation (see, among others, [2, 19, 20, 17, 7]), these techniques are still ad-hoc in that they are not formalized in the meta-tool: they happen "outside", and so all proofs and programs need to be written again and again as the language changes.

This paper describes a generic approach to modeling languages, by representing the definition of any given language as scrutable *data* instead of an ad-hoc *datatype*. This allows us to derive the datatype of terms (in varying levels of static guarantees), the implementation of renaming and substitution, and the proofs of these transformations behaving as expected, once and for all from this single description.

The full Agda implementation of the library described here is available at http://gergo.erdi.hu/projects/generic-syntax/.

## 2  Motivation

Suppose we want to use a type theory-based tool to study a typed language; in our running example, this language is going to be the simply typed lambda calculus[13]. The only types in this language are function types and a single base type:

```
data Ty : Set where
  •    : Ty
  _▷_  : Ty → Ty → Ty
```

The first goal is to give a representation of terms of our object language; in other words, a datatype whose values correspond to terms.

There is a spectrum of static semanics we could capture in this representation:

- The datatype could describe only the syntactical form of our language (using some type to represent variable names), with each constructor corresponding to an abstract grammar rule:

  ```
  data Form (name : Set) : Set where
    var : name                          → Form name
    lam : name → Form name              → Form name
    _·_ : Form name → Form name → Form name
  ```

- One step up, to represent well-scoped expressions, is to use de Bruijn indices for variables[8], and track the number of variables in scope at the type level[4]:

  ```
  data Expr (V : ℕ) : Set where
    var : Fin V              → Expr V
    lam : Expr (1 + V)       → Expr V
    _·_ : Expr V → Expr V → Expr V
  ```

  This allows us, for example, to talk about closed expressions:

  ```
  ClosedExpr : Set
  ClosedExpr = Expr 0

  close : ∀ {V} → Expr V → ClosedExpr
  close {zero}  e = e
  close {suc n} e = close (lam e)
  ```

- Finally, we can create a representation for well-typed terms, either by using the standard technique of typed pointers (into a context of types of variables in scope) to represent variables, and a datatype for terms indexed by the object-level type of said term[2] (an *intrinsically typed* representation):

  ```
  data Ctx : Set where
    ∅   : Ctx
    _,_ : Ctx → Ty → Ctx

  data Var : Ctx → Ty → Set where
    vz : ∀ {t}   {Γ}                   → Var (Γ , t) t
    vs : ∀ {u t} {Γ} (v : Var Γ t)     → Var (Γ , u) t

  infixl 23 _·_
  data Tm (Γ : Ctx) : Ty → Set where
    var : ∀ {t}   → Var Γ t                        → Tm Γ t
    lam : ∀ {u t} → Tm (Γ , t) u                   → Tm Γ (t ▷ u)
    _·_ : ∀ {u t} → Tm Γ (u ▷ t) → Tm Γ u → Tm Γ t
  ```

  Or, alternatively, using typed Parametric Higher-Order Syntax[12] representation:

  ```
  data HOTm (V : Ty → Set) : Ty → Set where
    var : ∀ {t}   → V t                            → HOTm V t
    lam : ∀ {u t} → (V t → HOTm V u)               → HOTm V (t ▷ u)
    _·_ : ∀ {u t} → HOTm V (u ▷ t) → HOTm V u → HOTm V t
  ```

Of course, there are various correspondences between these four representations – for example, we can erase the types from a Tm into an Expr, or resolve the names in a well-scoped Form to get an Expr using de Bruijn indices. We can write these, and more, correspondences as functions implementing mappings between these types; but there is nothing inherently tying together Form, Expr, Tm and HOTm since they are four independent datatypes as far as our metalanguage Agda is concerned.

Moreover, even once we write these functions, we are usually not done yet by this point. Languages with variables have functorial (renaming) and monadic (substitution) structure that is needed for the syntactic accounting of most reduction rules. One approach is to represent renamings and simultaneous substitutions as functions from variables:

```
_→̇_ : ∀ {a b c} {A : Set a} → (A → Set b) → (A → Set c) → Set _
F →̇ G = ∀ {t} → F t → G t

Ren : Ctx → Ctx → Set
Ren Γ Δ = Var Δ →̇ Var Γ

map : ∀ {Γ Δ} → Ren Γ Δ → Tm Δ →̇ Tm Γ

Sub : Ctx → Ctx → Set
Sub Γ Δ = Var Δ →̇ Tm Γ

bind : ∀ {Γ Δ} → Sub Γ Δ → Tm Δ →̇ Tm Γ
```

Another approach is to use some data structures to describe renamings and simultaneous substitutions. This has the benefit that equality of the *contents* of the renaming/substitution automatically lifts to extensional equality of their *action*. Here, and in the rest of this paper, we are going to use, and reserve the more evocative names ren and sub for, this second approach, following [10]: representing renamings as order-preserving embeddings of contexts, and substitutions as a snoc-list of terms:

```
infix 3 _⊇_
data _⊇_ : Ctx → Ctx → Set where
  done : ∅ ⊇ ∅
  drop : ∀ {t Γ Δ} → Γ ⊇ Δ → Γ , t ⊇ Δ
  keep : ∀ {t Γ Δ} → Γ ⊇ Δ → Γ , t ⊇ Δ , t

ren–var : ∀ {Γ Δ} → Δ ⊇ Γ → Var Γ →̇ Var Δ

data Env (A : Ty → Set) : Ctx → Set where
  ∅ : Env A ∅
  _,_ : ∀ {t Δ} → (σ : Env A Δ) → (e : A t) → Env A (Δ , t)

lookup : ∀ {A Δ} → Env A Δ → Var Δ →̇ A

_⊢*_ : Ctx → Ctx → Set
Γ ⊢* Δ = Env (Tm Γ) Δ
```

```
sub−var : ∀ {Γ Δ} → Γ ⊢* Δ → Var Δ →̇ Tm Γ
sub−var = lookup

ren : ∀ {Γ Δ} → Γ ⊇ Δ → Tm Δ →̇ Tm Γ
ren ρ (var v) = var (ren−var ρ v)
ren ρ (lam e) = lam (ren (keep ρ) e)
ren ρ (f · e) = ren ρ f · ren ρ e

sub : ∀ {Γ Δ} → Γ ⊢* Δ → Tm Δ →̇ Tm Γ
sub σ (var v) = sub−var σ v
sub σ (lam e) = lam (sub (shift σ) e)
sub σ (f · e) = sub σ f · sub σ e
```

We omit the definition of some helper functions here, because there is nothing interesting about them. In fact, it would be fair to say there is nothing interesting in the implementation of map/ren and bind/sub either; however, they are included here so we can note that they are all defined in terms of pattern matching on the constructors of Tm. This means not only that the implementation of these two *functions* need to be changed every time we wish to change our object language; but also, all *proofs about their behaviour* as well; proofs such as the following:

```
refl_r : ∀ {Γ} → Γ ⊇ Γ
refl_r {∅}     = done
refl_r {Γ , _} = keep refl_r

ren−refl : ∀ {Γ t} (e : Tm Γ t) → ren refl_r e ≡ e
ren−refl (var v) rewrite ren−var−refl v = refl
ren−refl (lam e) rewrite ren−refl e = refl
ren−refl (f · e) rewrite ren−refl f | ren−refl e = refl
```

Just like the definition of ren, arguably there is nothing interesting about the proof of ren−refl and similar theorems. However, for a lot of applications, they still need to be written, by hand, and updated whenever the definition of Tm changes. In fact, the key motivation for this paper was encountering an Agda library[16] implementing the usual renaming and substitution lemmas that starts with the following comment:

```
-- Copy/paste everything from below,
-- then add/remove cases as necessary,
-- depending on the definition of the syntax.
```

How can we turn these functions and proofs into a reusable library?

## 3 Generic description of syntaxes

The standard approach[3, 18, 21, 5] for writing *type-generic programs* in a dependently-typed setting is to create a datatype that can be used to point into the universe of types. Generic functions and proofs can then be written by recursing, simultaneously, on a code of a type and values of the type that is the interpretation of the given code. If needed, the coded representation can also be chosen to only encode a circumscribed sub-universe of all types.

Similarly, the road to writing functions like type erasure and simultaneous substitution, and proofs like renaming by reflexivity, once and for all, is paved with writing them not in terms of a bespoke datatype for one given object language. Instead, we create in a library a way of writing a description of a language, and then, given a particular description by the user, synthesize typed and untyped representations, functions between these representations, and proofs about these functions in the form of *syntax-generic programs*.

Assuming we have $Ty$ : Set provided by the user, the central datatype for describing languages is Desc defined below:

```
data Binder : Set where
  bound unbound : Binder

Shape : ℕ → ℕ → Set
Shape n k = Vec (Vec Binder n) k

data Desc : Set₁ where
  sg   : (A : Set) (k : A → Desc) → Desc
  node : (n : ℕ) {k : ℕ} (shape : Shape n k) (wt : Vec Ty n → Vec Ty k → Ty → Set) → Desc
```

The sg constructor is the way to put arbitrary user-specified data into the syntax nodes. Because the rest of the description depends on the stored piece of data, this is also the mechanism by which the alternatives of the object grammar can be encoded. This matches the corresponding behaviour of [1].

The idea behind the node constructor will become clear when we get to deriving a typed representation: the goal is to enable a global view of all newly-bound variables and the types of all sub-terms. $n$ is the number of newly bound variables, and $k$ is the number of subterms; *shape* describes the subterm structure: which newly bound variable is in scope in which subterm. $wt$ is the user-supplied well-typedness constraint: a proposition over the types of the $n$ newly-bound variables, the types of the $k$ sub-terms, and the type of the term itself.

To give a feel for Descriptions, let's look at some examples, in increasing order of complexity. For function APPlication, we introduce no new variables, so the shape just prescribes two subterms. The well-typedness constraint ensures the type of the function matches the type of the argument and the result.

```
APP : Desc
APP = node 0 ([] ∷ [] ∷ [])
  λ { [] (t₁ ∷ t₂ ∷ []) t₀ → t₁ ≡ t₂ ▷ t₀ }
```

LAMbda abstractions can be added either in Curry or Church style. Both bind one new variable and has one subterm (in which the newly bound variable is in scope). The well-typedness constraint connects the type of the newly bound variable $t'$ and the type of the subterm $u$ to the type of the whole lambda abstraction $t_0$. In Church style, the parameter type is stored in a sg node and also used in the typing constraint:

```
data Style : Set where
  Curry Church : Style

LAM_style : Style → Desc
LAM Curry style = node 1 ((bound :: []) :: [])
  λ { (t' :: []) (u :: []) t₀ → t₀ ≡ t' ▷ u }
LAM Church style = sg Ty λ t → node 1 ((bound :: []) :: [])
  λ { (t' :: []) (u :: []) t₀ → t ≡ t' × t₀ ≡ t ▷ u }
```

For (non-recursive) LET, there are two subterms for the one newly bound variable: the first one is the definition, and the second one is the rest of the term where the new variable is in scope. Of course, the type of the definition $u'$ must match the type of the newly introduced variable $u$; additionally, the type $t_0$ of the whole term is the same as the type $t$ of term where the new variable is in scope:

```
LET : Desc
LET = node 1 ((unbound :: []) :: (bound :: []) :: [])
  λ { (u :: []) (u' :: t' :: []) t₀ → u ≡ u' × t₀ ≡ t' }
```

The only tweak we need to make for the recursive version (i.e. LETREC) is to make the newly bound variable visible in the definition as well:

```
LETREC : Desc
LETREC = node 1 ((bound :: []) :: (bound :: []) :: [])
  λ { (u :: []) (u' :: t' :: []) t₀ → u ≡ u' × t₀ ≡ t' }
```

## 4 Representations for terms

Now that we have a way of describing our language, it is time to start reaping benefits from it. First, to do anything meaningful with terms, we need a way to represent them, i.e. some Agda type (for a given Description) whose (Agda) values correspond to terms of the object language.

### 4.1 Form: Syntax only

For the representation of purely syntactic Forms with explicit names (with no scoping rules enforced), we simply take the arity of new names and the arity of subterms from the $n$ and $k$ arguments of node. Crucially, and in contrast to [1], because there is no Desc continuation in node that would depend on any typing information, we can just ignore *wt* and *shape*:

```
module Unscoped (desc : Desc) (name : Set) where
  mutual
    data Form : Set where
      var : name      → Form
      con : Con desc  → Form

    data Con : Desc → Set where
      sg    : ∀ {A k} x → Con (k x) → Con (sg A k)
      node  : ∀ {n k sh wt} (ns : Vec name n) (es : Children k)
```

G. Érdi7```
    → Con (node n {k} sh wt)

Children : ℕ → Set
Children = Vec Form
```

## 4.2 Expr: Untyped, well-scoped

The untyped representation is still indexed by the number of variables in scope. At any inductive site, we increase that by the number of newly bound variables that are in scope, as prescribed by the *shape* in the node constructor of Desc. Because this representation is untyped, the well-typedness constraint *wt* is not used.

```
countV : ∀ {n} → Vec Binder n → ℕ
countV []              = 0
countV (bound :: bs)   = suc (countV bs)
countV (unbound :: bs) = countV bs

module Untyped (desc : Desc) where
  mutual
    data Expr (V : ℕ) : Set where
      var : Fin V        → Expr V
      con : Con V desc   → Expr V

    data Con (V : ℕ) : Desc → Set where
      sg    : ∀ {A k} x → Con V (k x) → Con V (sg A k)
      node  : ∀ {n k sh wt} (es : Children V sh)
            → Con V (node n {k} sh wt)

    Children : ∀ {n k} → ℕ → Shape n k → Set
    Children V = All (λ bs → Expr (countV bs + V))
```

## 4.3 Tm: Intrinsically well-typed

First, to compute the extended context of subterms in the generic case when there are multiple newly bound variables, we make an object type-ornamented version of _+_ (à la [20]) and countV:

```
_<><_ : Ctx → List Ty → Ctx
Γ <>< [] = Γ
Γ <>< (t :: ts) = (Γ , t) <>< ts
```

For the intrinsically typed representation, we generalize the standard technique of representing GADTs[23]: the indices corresponding to the types of the newly bound variables and the subterms are existentially bound, and the index of the term itself is turned into a parameter; then, an explicit witness of well-typedness restores guardedness. This is a generalization in the sense that well-typedness can be an arbitrary proposition, not necessarily just a conjunction of equalities.



```
visible : ∀ {n} → Vec Binder n → Vec Ty n → List Ty
visible []            []        = []
visible (bound :: bs)   (t :: ts) = t :: visible bs ts
visible (unbound :: bs) (_ :: ts) = visible bs ts

module Typed (desc : Desc) where
  mutual
    data Tm (Γ : Ctx) : Ty → Set where
      var : ∀ {t} → Var Γ t      → Tm Γ t
      con : ∀ {t} → Con Γ t desc → Tm Γ t

    data Con (Γ : Ctx) (t : Ty) : Desc → Set where
      sg   : ∀ {A k} x → Con Γ t (k x) → Con Γ t (sg A k)
      node : ∀ {n k sh wt} (ts₀ : Vec Ty n) {ts : Vec Ty k}
        (es : Children Γ ts₀ sh ts)
        {{_ : wt ts₀ ts t}}
        → Con Γ t (node n sh wt)

    Children : ∀ {n k} → Ctx → Vec Ty n → Shape n k → Vec Ty k → Set
    Children Γ ts₀ = All₂ (λ bs → Tm (Γ <>< visible bs ts₀))
```

### 4.4 HOTm: Typed PHOAS

Although the PHOAS representation is not used in the rest of this paper, we included it here for the sake of completeness.[1] It differs from Tm only in that Children subterms are represented as functions from *V*ariables to (higher-order) terms:

```
module PHOAS (desc : Desc) where
  mutual
    data HOTm (V : Ty → Set) : Ty → Set where
      var : ∀ {t} → V t        → HOTm V t
      con : ∀ {t} → Con V t desc → HOTm V t

    data Con (V : Ty → Set) (t : Ty) : Desc → Set where
      sg : ∀ {A k} x → Con V t (k x) → Con V t (sg A k)
      node : ∀ {n k sh wt} (ts₀ : Vec Ty n) {ts : Vec Ty k}
        (es : Children V ts₀ sh ts)
        {{_ : wt ts₀ ts t}}
        → Con V t (node n sh wt)

    Children : ∀ {n k} → (Ty → Set) → Vec Ty n → Shape n k → Vec Ty k → Set
    Children V ts₀ = All₂ (λ bs t → (All V (visible bs ts₀) → HOTm V t))
```

---

[1] Our full library implementation also includes a conversion function from Tm to HOTm



### 4.5 Example: Type erasure

The building blocks introduced so far already allow us to write some interesting functions generically for all syntaxes. As an example, here is a conversion function between the typed and the untyped representation of terms. Because the two representations are computed from a single shared Description, the type of untype is more valuable as a specification than it would be if it converted between two ad-hoc datatypes.

```
untype−var : ∀ {Γ t} → Var Γ t → Fin (size Γ)
untype−var vz = zero
untype−var (vs v) = suc (untype−var v)

untype : ∀ {Γ t} → Tm Γ t → Expr (size Γ)
untype (var v) = var (untype−var v)
untype (con e) = con (untype−con e)
  where
    untype* : ∀ {Γ n k sh ts₀ ts} → T.Children {n} {k} Γ ts₀ sh ts → U.Children (size Γ) sh
    untype*                    []          = []
    untype* {sh = bs :: _} (e :: es) = subst Expr (size−<>< _ bs _) (untype e) :: untype* es

    untype−con : ∀ {Γ t c} → T.Con Γ t c → U.Con (size Γ) c
    untype−con (sg x e)      = sg x (untype−con e)
    untype−con (node ts₀ es) = node (untype* es)
```

## 5 Type-preserving renaming and substitution

Now that we have a typed representation, implementing type-preserving renaming and substitution becomes a straightforward matter of recursing simultaneously on the description (implicitly) and a given term of the language corresponding to that description (explicitly):

```
mutual
  ren : ∀ {Γ Δ} → Γ ⊇ Δ → Tm Δ →̇ Tm Γ
  ren ρ (var v) = var (ren−var ρ v)
  ren ρ (con c) = con (ren−con ρ c)

  ren−con : ∀ {d Γ Δ t} → Γ ⊇ Δ → Con Δ t d → Con Γ t d
  ren−con ρ (sg x k)       = sg x (ren−con ρ k)
  ren−con ρ (node ts₀ es) = node ts₀ (ren* ρ es)

  ren* : ∀ {Γ Δ} {n k ts₀ sh ts} → Γ ⊇ Δ → Children {n} {k} Δ ts₀ sh ts → Children Γ ts₀ sh ts
  ren*              ρ []         = []
  ren* {sh = bs :: _} ρ (e :: es) = ren (keep* (visible bs _) ρ) e :: ren* ρ es

mutual
  sub : ∀ {Γ Δ} → Γ ⊢* Δ → Tm Δ →̇ Tm Γ
  sub σ (var v) = sub−var σ v
  sub σ (con c) = con (sub−con σ c)
```



```
sub−con : ∀ {Γ Δ t c} → Γ ⊢* Δ → Con Δ t c → Con Γ t c
sub−con σ (sg x c)      = sg x (sub−con σ c)
sub−con σ (node ts₀ es) = node ts₀ (sub* σ es)

sub* : ∀ {Γ Δ n k sh ts₀ ts} → Γ ⊢* Δ → Children {n} {k} Δ ts₀ sh ts → Children Γ ts₀ sh ts
sub*                σ []       = []
sub* {sh = bs :: _} σ (e :: es) = sub (shift* (visible bs _) σ) e :: sub* σ es
```

Finally, we can fulfill our original goal of proving properties of renamings and substitutions once and for all. Our library implementation contains all proofs from [16] adapted to use our generic representation, thus avoiding the need to change proofs as the object language is changed. As an illustration, here is ren−refl from earlier, generically for all syntaxes:

```
mutual
  ren−refl : ∀ {Γ t} → (e : Tm Γ t) → ren reflᵣ e ≡ e
  ren−refl (var v) rewrite ren−var−refl v = refl
  ren−refl (con e) rewrite ren−con−refl e = refl

  ren−con−refl : ∀ {c Γ t} (e : Con Γ t c) → ren−con reflᵣ e ≡ e
  ren−con−refl (sg x e)      rewrite ren−con−refl e = refl
  ren−con−refl (node ts es)  rewrite ren*−refl es = refl

  ren*−refl : ∀ {Γ n k sh ts₀ ts} (es : Children {n} {k} Γ ts₀ sh ts) → ren* reflᵣ es ≡ es
  ren*−refl                              []        = refl
  ren*−refl {Γ} {sh = bs :: _} {ts₀} (e :: es) rewrite keep*−refl {Γ} (visible bs ts₀) =
    cong₂ _::_ (ren−refl e) (ren*−refl es)
```

## 6  Simply Typed Lambda Calculus, generically

In this section, we show details of modeling the simply typed lambda calculus by giving its Description and using the derived representations for terms.

Using the definitions from section 3, we put it all together using a simple datatype for tagging nodes by their grammar production rules:

```
data `STLC : Set where
  `app `lam : `STLC

STLC : Desc
STLC = sg `STLC λ
  { `app → APP
  ; `lam → LAM Curry style
  }
```

Note that there is no explicit mention of variable occurrences (the var constructor of our original Tm datatype in section 2) in this description. That is because in the framework we are presenting, variables are understood to be (unavoidably) part of the syntax. In fact, the concept of variables is *the* crucial



differentiator of the syntaxes we consider here, versus arbitrary inductive definitions. This is because the functorial and monadic structure we expect of terms is defined in terms of variables.

For the various datatypes representing terms, we can recover _·_ and lam (the constructors of the datatypes from section 2) with some pattern synonyms[27].[2] For the Typed representation, the pattern synonyms include matching on the explicit well-typedness witnesses to refine the scrutinee Tm's type index whenever they are used; for the Untyped and Unscoped representations, the pattern synonyms contain less detail while matching on the same shape:

```
open Untyped STLC

-- _·_ : ∀ {V} → Expr V → Expr V → Expr V
pattern _·_ f e = con (sg `app (node (f ∷ e ∷ [])))

-- lam : ∀ {V} → Ty → Expr (1 + V) → Expr V
pattern lam t e = con (sg `lam (sg t (node (e ∷ []))))

open Unscoped STLC Name

-- _·_ : Form → Form → Form
pattern _·_ f e = con (sg `app (node [] (f ∷ e ∷ [])))

-- lam : Name → Ty → Form → Form
pattern lam n t e = con (sg `lam (sg t (node (n ∷ []) (e ∷ []))))

open Typed STLC

-- _·_ : ∀ {Γ t u} → Tm Γ (t ▸ u) → Tm Γ t → Tm Γ u
pattern _·_ f e = con (sg `app (node [] (f ∷ e ∷ []) {{refl}}))

-- lam : ∀ {Γ t u} → Tm (Γ , t) u → Tm Γ (t ▸ u)
pattern lam e = con (sg `lam (node (_ ∷ []) (e ∷ []) {{refl}}))
```

## 6.1 Example: Desugaring let bindings

The following example demonstrates that once we define these handy pattern synonyms, writing non-generic intensional transformations is no more unwieldy than with a direct, datatype-based approach.

We define, in one fell swoop, languages in two Flavours in Church and Curry style, with `let bindings only permitted in the sugared flavour. (let is unfortunately an Agda keyword, hence the name letvar for the pattern synonym)

```
data Flavour : Set where
  sugared desugared : Flavour

data `STLC : Flavour → Set where
```

---

[2]Since Agda's pattern synonyms are not typed, there is no way to attach a type signature to them[26]. In this paper, we will give signatures in comments just for clarity.



```
  `app `lam : ∀ {φ} → `STLC φ
  `let : `STLC sugared

STLC : Flavour → Style → Desc
STLC φ s = sg (`STLC φ) λ
  { `app → APP
  ; `lam → LAM s style
  ; `let  → LET
  }

Tm : Flavour → Style → Ctx → Ty → Set
Tm φ s = Typed.Tm (STLC φ s)

pattern _·_ f e    = con (sg `app (node [] (f :: e :: []) {{refl}}))
pattern lam e      = con (sg `lam (node (_ :: []) (e :: []) {{refl}}))
pattern lam′ t e   = con (sg `lam (sg t ((node (_ :: []) (e :: []) {{refl , refl}})))))
pattern letvar e₀ e = con (sg `let (node (_ :: []) (e₀ :: e :: []) {{refl , refl}}))
```

Then we can define a desugaring transformation in exactly the same way as we would do in direct style; the type ensures that the function is object type-preserving, and the result contains no let bindings. The transformation of both Curry and Church style languages is implemented in a single function.

```
desugar : ∀ {s φ Γ} → Tm φ s Γ →̇ Tm desugared s Γ
desugar              (var v)       = var v
desugar              (f · e)       = desugar f · desugar e
desugar {Curry}  (lam e)       = lam (desugar e)
desugar {Church} (lam′ t e)    = lam′ t (desugar e)
desugar {Curry}  (letvar e₀ e) = lam (desugar e) · desugar e₀
desugar {Church} (letvar e₀ e) = lam′ _ (desugar e) · desugar e₀
```

## 6.2 Example: Normalization

We have formalized a normalization proof of STLC based on the approach presented in [25]. Since our terms are intrinsically typed, and the simultaneous substitution function is type-preserving, we can avoid a lot of otherwise lengthy type preservation proofs. We show our formalization in broad strokes here to highlight the use of the generic syntax library.

Small-step semantics of STLC is defined with a type-preserving reduction relation between typed terms. sub is crucially used in defining the reduction rule for applying a lambda abstraction.

```
data Value : ∀ {Γ t} → Tm Γ t → Set where
  ↓lam : ∀ {Γ t u} → (e : Tm (Γ , u) t) → Value {Γ} {u ▷ t} (lam e)

infix 19 _==>_ _==>*_

data _==>_ {Γ} : ∀ {t} → Rel (Tm Γ t) lzero where
  app–lam : ∀ {t u} {e : Tm _ u} → (f : Tm _ t) → Value e         → lam f · e ==> sub (refl , e) f
  app–fun : ∀ {t u} {f f′ : Tm _ (u ▷ t)} → f ==> f′ → ∀ e        → f · e ==> f′ · e
```



app−arg : ∀ {t u} {f : Tm _ (u ▷ t)} {e e'} → Value f → e ==> e' → f · e ==> f · e'

_==>*_ : ∀ {Γ t} → Rel (Tm Γ t) _
_==>*_ = Star _==>_

The bulk of the proof proceeds exactly as in [25] towards the final goal of

normalization : ∀ {t} → (e : Tm ∅ t) → Halts e

At one point in the course of proving normalization, we arrive at a proof obligation for

app−sub : lam (sub (shift σ) f) · e ==>* sub (σ , e) f

which can be filled in by delegating most of the heavy lifting to our library of proofs; namely the following three:

sub−⊢*⊇ : (σ : Γ ⊢* Θ) (ρ : Θ ⊇ Δ)        (e : Tm Δ t) → sub (σ ⊢*⊇ ρ) e ≡ sub σ (ren ρ e)
sub−refl :                                   (e : Tm Γ t) → sub refl e ≡ e
sub−⊢⊢* : (σ₂ : Γ ⊢* Θ) (σ₁ : Θ ⊢* Δ) (e : Tm Δ t) → sub (σ₂ ⊢⊢* σ₁) e ≡ sub σ₂ (sub σ₁ e)

Extending our language with a new boolean type and the corresponding eliminator for conditional expressions involved changing the Description to

```
data `STLC : Set where
  `app `lam `true `false `if : `STLC

STLC : Desc
STLC = sg `STLC λ
  { `app   → node 0 ([] :: [] :: [])        λ { []      (t₁ :: t₂ :: [])   t₀ → t₁ ≡ t₂ ▷ t₀ }
  ; `lam   → node 1 ((bound :: []) :: [])  λ { (t :: []) (u :: [])          t₀ → t₀ ≡ t ▷ u }
  ; `true  → node 0 []                      λ { []      []                  t₀ → t₀ ≡ bool }
  ; `false → node 0 []                      λ { []      []                  t₀ → t₀ ≡ bool }
  ; `if    → node 0 ([] :: [] :: [] :: []) λ { []      (b :: t₁ :: t₂ :: []) t₀ → b ≡ bool × t₁ ≡ t₀ × t₂ ≡ t₀ }
  }
```

and adding new axioms to the Value and ==> propositions:

```
data Value : ∀ {Γ t} → Tm Γ t → Set where
  ↓lam   : ∀ {Γ t u} → (e : Tm (Γ , u) t) → Value {Γ} {u ▷ t} (lam e)
  ↓true  : ∀ {Γ} → Value {Γ} true
  ↓false : ∀ {Γ} → Value {Γ} false

data _==>_ {Γ} : ∀ {t} → Rel (Tm Γ t) lzero where
  app−lam : ∀ {t u} {e : Tm _ u} → (f : Tm _ t) → Value e            → lam f · e ==> sub (refl , e) f
  app−fun : ∀ {t u} {f f' : Tm _ (u ▷ t)} → f ==> f' → ∀ e           → f · e ==> f' · e
  app−arg : ∀ {t u} {f : Tm _ (u ▷ t)} {e e'} → Value f → e ==> e'   → f · e ==> f · e'
  if−cond : ∀ {t} {b b'} → b ==> b' → (thn els : Tm _ t)              → if b thn els ==> if b' thn els
  if−true :   ∀ {t} → (thn els : Tm _ t)                              → if true thn els ==> thn
  if−false :  ∀ {t} → (thn els : Tm _ t)                              → if false thn els ==> els
```

but, upholding the promise of a generic library for manipulating terms, none of the proofs involving syntactic transformations, like app−sub or its building blocks, needed any updating.



# 7 Conclusions and future work

In this paper, we have described the datatype Desc that can be used to give a machine-readable definition of a statically typed language with variable-binding structures. We have implemented a library in the dependently typed programming language Agda that provides untyped, well-scoped, and well-typed representations of terms; type-preserving renaming and substitution; and proofs of their well-behaving.

We have demonstrated the use of the library by formalizing a normalization proof of the simply-typed lambda calculus; and then observed the ease with which the object language can be changed or extended without labouriously rewriting the syntactic transformations and proofs.

We have also recovered object language-specific functions like let-desugaring and (in the full version of the library) typechecking that are implemented by pattern matching on pattern synonyms that exhibit the correspondence between the description-derived representation and the traditional standard approach.

One possible direction for future work is support for languages with multiple binding namespaces by multiplexing the context. These contexts can be orthogonal, like the truth and validity hypotheses in modal logic *à la* [24], where variable occurrences would be indexed by which namespace they point into; this index can then be used by the well-typedness constraint to enforce restrictions on occurrence. Or the contexts could correspond to multiple sorts and be nested, for polymorphic languages like System F[14]: terms have types have kinds, and the types of term variables available in a given context can refer to type variables. This requires substitution to be able to shift on multiple axes at the same time, as discussed in [15].

Since well-typedness is represented as an arbitrary proposition over subterm types, bound variable types, and the type of the term itself, it can be any relation between them, not just a conjunction of equalities. We have not explored this avenue yet, but it seems applicable to supporting languages with subtyping[9], or those that use unification in their type system[11] (and not just in implementing a typechecker). However, treating renamings and substitutions as *exact type*-preserving may be inadequate in these settings.

# References


[1] Guillaume Allais, Robert Atkey, James Chapman, Conor McBride & James McKinna (2018): *A Scope-and-Type Safe Universe of Syntaxes with Binding, Their Semantics and Proofs*. Available at https://github.com/gallais/generic-syntax.

[2] Guillaume Allais, James Chapman, Conor McBride & James McKinna (2017): *Type-and-scope safe programs and their proofs*. In: *Proceedings of the 6th ACM SIGPLAN Conference on Certified Programs and Proofs*, ACM, pp. 195–207.

[3] Thorsten Altenkirch & Conor McBride (2003): *Generic programming within dependently typed programming*. In: *Generic Programming*, Springer, pp. 1–20.

[4] Thorsten Altenkirch & Bernhard Reus (1999): *Monadic presentations of lambda terms using generalized inductive types*. In: *International Workshop on Computer Science Logic*, Springer, pp. 453–468.

[5] Stevan Andjelkovic (2011): *A family of universes for generic programming*.

[6] Bruno Barras, Samuel Boutin, Cristina Cornes, Judicaël Courant, Jean-Christophe Filliatre, Eduardo Gimenez, Hugo Herbelin, Gerard Huet, Cesar Munoz, Chetan Murthy et al. (1997): *The Coq proof assistant reference manual*. Ph.D. thesis, Inria.

[7] Nick Benton, Chung-Kil Hur, Andrew J Kennedy & Conor McBride (2012): *Strongly typed term representations in Coq*. Journal of automated reasoning 49(2), pp. 141–159.






[8] Richard S Bird & Ross Paterson (1999): *De Bruijn notation as a nested datatype*. *Journal of functional programming* 9(1), pp. 77–91.

[9] Luca Cardelli, Simone Martini, John C Mitchell & Andre Scedrov (1994): *An extension of system F with subtyping*. *Information and Computation* 109(1-2), pp. 4–56.

[10] James Maitland Chapman (2009): *Type checking and normalisation*. Ph.D. thesis, University of Nottingham.

[11] Olaf Chitil (2001): *Compositional explanation of types and algorithmic debugging of type errors*. In: *ACM SIGPLAN Notices*, 36, ACM, pp. 193–204.

[12] Adam Chlipala (2008): *Parametric higher-order abstract syntax for mechanized semantics*. In: *ACM Sigplan Notices*, 43, ACM, pp. 143–156.

[13] Alonzo Church (1940): *A formulation of the simple theory of types*. *The journal of symbolic logic* 5(2), pp. 56–68.

[14] Jean-Yves Girard (1971): *Une Extension De L'Interpretation De Gödel a L'Analyse, Et Son Application a L'Elimination Des Coupures Dans L'Analyse Et La Theorie Des Types*. In: *Studies in Logic and the Foundations of Mathematics*, 63, Elsevier, pp. 63–92.

[15] Jonas Kaiser, Steven Schäfer & Kathrin Stark (2017): *Autosubst 2: Towards Reasoning with Multi-Sorted de Bruijn Terms and Vector Substitutions*. In: *Proceedings of the Workshop on Logical Frameworks and Meta-Languages: Theory and Practice*, ACM, pp. 10–14.

[16] András Kovács (2017): *Example of complete STLC substitution calculus*. https://gist.github.com/AndrasKovacs/bd6a6333e4eecd7acb0eb9d98f7e0cb8.

[17] Daniel R Licata & Robert Harper (2009): *A universe of binding and computation*. In: *ACM Sigplan Notices*, 44, ACM, pp. 123–134.

[18] Per Martin-Löf & Giovanni Sambin (1984): *Intuitionistic type theory*. 9, Bibliopolis Napoli.

[19] Conor McBride (2005): *Type-preserving renaming and substitution*.

[20] Conor McBride (2013): *Dependently Typed Metaprogramming (in Agda)*.

[21] Bengt Nordström, Kent Petersson & Jan M Smith (1990): *Programming in Martin-Löf's type theory*. 200, Oxford University Press Oxford.

[22] Ulf Norell (2008): *Dependently typed programming in Agda*. In: *International School on Advanced Functional Programming*, Springer, pp. 230–266.

[23] Simon Peyton Jones, Dimitrios Vytiniotis, Stephanie Weirich & Geoffrey Washburn (2006): *Simple unification-based type inference for GADTs*. In: *ACM SIGPLAN Notices*, 41, ACM, pp. 50–61.

[24] Frank Pfenning & Rowan Davies (2001): *A judgmental reconstruction of modal logic*. *Mathematical structures in computer science* 11(4), pp. 511–540.

[25] Benjamin C. Pierce, Arthur Azevedo de Amorim, Chris Casinghino, Marco Gaboardi, Michael Greenberg, Cătălin Hriţcu, Vilhelm Sjöberg & Brent Yorgey (2017): *Software Foundations*. Electronic textbook. Version 5.0. http://www.cis.upenn.edu/~bcpierce/sf.

[26] The Agda Team (2016): *Release notes for Agda 2 version 2.3.2*. http://wiki.portal.chalmers.se/agda/pmwiki.php?n=Main.Version-2-3-2. Accessed: 2017-12-30.

[27] Mark Tullsen (2000): *First Class Patterns?* In: *International Symposium on Practical Aspects of Declarative Languages*, Springer, pp. 1–15.